\documentclass[conference]{IEEEtran}
\IEEEoverridecommandlockouts
\usepackage{cite}
\usepackage{amsmath,amssymb,amsfonts}
\usepackage{algorithmic}
\usepackage{graphicx}
\usepackage{textcomp}
\usepackage{xcolor}
\usepackage{url}
\usepackage{mdframed}
\usepackage{csquotes}

\def\BibTeX{{\rm B\kern-.05em{\sc i\kern-.025em b}\kern-.08em
    T\kern-.1667em\lower.7ex\hbox{E}\kern-.125emX}}
\begin{document}

\title{Machine Learning-based xApp for Dynamic Resource Allocation in O-RAN Networks}
\author{\IEEEauthorblockN{Mohammed M. H. Qazzaz$^{1,2}$\qquad Łukasz Kułacz$^{3,4}$\qquad Adrian Kliks$^{3,4}$\\\qquad Syed A. Zaidi$^{1}$\qquad Marcin Dryjanski$^{4}$\qquad Des McLernon$^{1}$}
\IEEEauthorblockA{\textit{$^{1}$ School of Electronic and Electrical Engineering,}
\textit{University of Leeds,} Leeds, UK \\
\textit{$^{2}$ College of Electronics Engineering,}
\textit{Ninevah University,} Mosul, Iraq \\
\textit{$^{3}$ Institute of Radiocommunications, Poznan University of Technology,} Poznan, Poland \\
\textit{$^{4}$ Rimedo Labs,} Poznan, Poland \\
Corresponding Author: Mohammed M. H. Qazzaz (ml14mmh@leeds.ac.uk)}
}

\maketitle

\begin{abstract}
The disaggregated, distributed and virtualised implementation of radio access networks allows for dynamic resource allocation. These attributes can be realised by virtue of the Open Radio Access Networks (O-RAN) architecture. In this article, we tackle the issue of dynamic resource allocation using a data-driven approach by employing Machine Learning (ML). We present an xApp-based implementation for the proposed ML algorithm. The core aim of this work is to optimise resource allocation and fulfil Service Level Specifications (SLS). This is accomplished by dynamically adjusting the allocation of Physical Resource Blocks (PRBs) based on traffic demand and Quality of Service (QoS) requirements. The proposed ML model effectively selects the best allocation policy for each base station and enhances the performance of scheduler functionality in O-RAN - Distributed Unit (O-DU). We show that an xApp implementing the Random Forest Classifier can yield high (85\%) performance accuracy for optimal policy selection. This can be attained using the O-RAN instance state input parameters over a short training duration. 

\end{abstract}


\section{Introduction}\label{introduction}
\subsection{Motivation}
The traditional cellular network model, relying upon a single vendor's proprietary hardware and software solutions, limits flexibility, increases costs and hampers innovation \cite{bonati2020open}. Open RAN (O-RAN), also known as Open Radio Access Network, has revolutionised the telecommunications industry by decoupling hardware and software components, fostering an open and interoperable ecosystem. This transformative approach aims to mitigate traditional network limits by enhancing network adaptability, reducing costs, and promoting innovation. Additionally, the integration of artificial intelligence (AI) and machine learning (ML) capabilities within O-RAN holds tremendous promise for optimising network performance, automating operations, and enabling intelligent decision-making \cite{masur2022artificial,plantin2021geopolitical}.

In that context, traditionally regarded as monolithic and immutable \enquote{black-box} systems, cellular networks are transitioning towards more flexible, software-based open architectures following the O-RAN paradigm. This paradigm promotes openness, virtualisation, and programmability of RAN functionalities and components, enabling data-driven intelligent control loops for cellular systems \cite{wypior2022open}. O-RAN empowers network operators to support bespoke services on shared physical infrastructures and dynamically reconfigure them based on network conditions and user demand. The O-RAN Alliance, a notable standardisation body, develops specifications to apply O-RAN principles to prevailing radio access technologies, including 3rd Generation Partnership Project (3GPP) LTE and 5G networks. O-RAN encompasses RAN Intelligent Controllers (RICs) that operate at different timescales, enabling data-driven applications. These applications optimise network performance by leveraging live data received from the RAN through standardised and open interfaces \cite{bonati2021intelligence}.

\begin{figure}
  \centering
  \includegraphics[width=\linewidth]{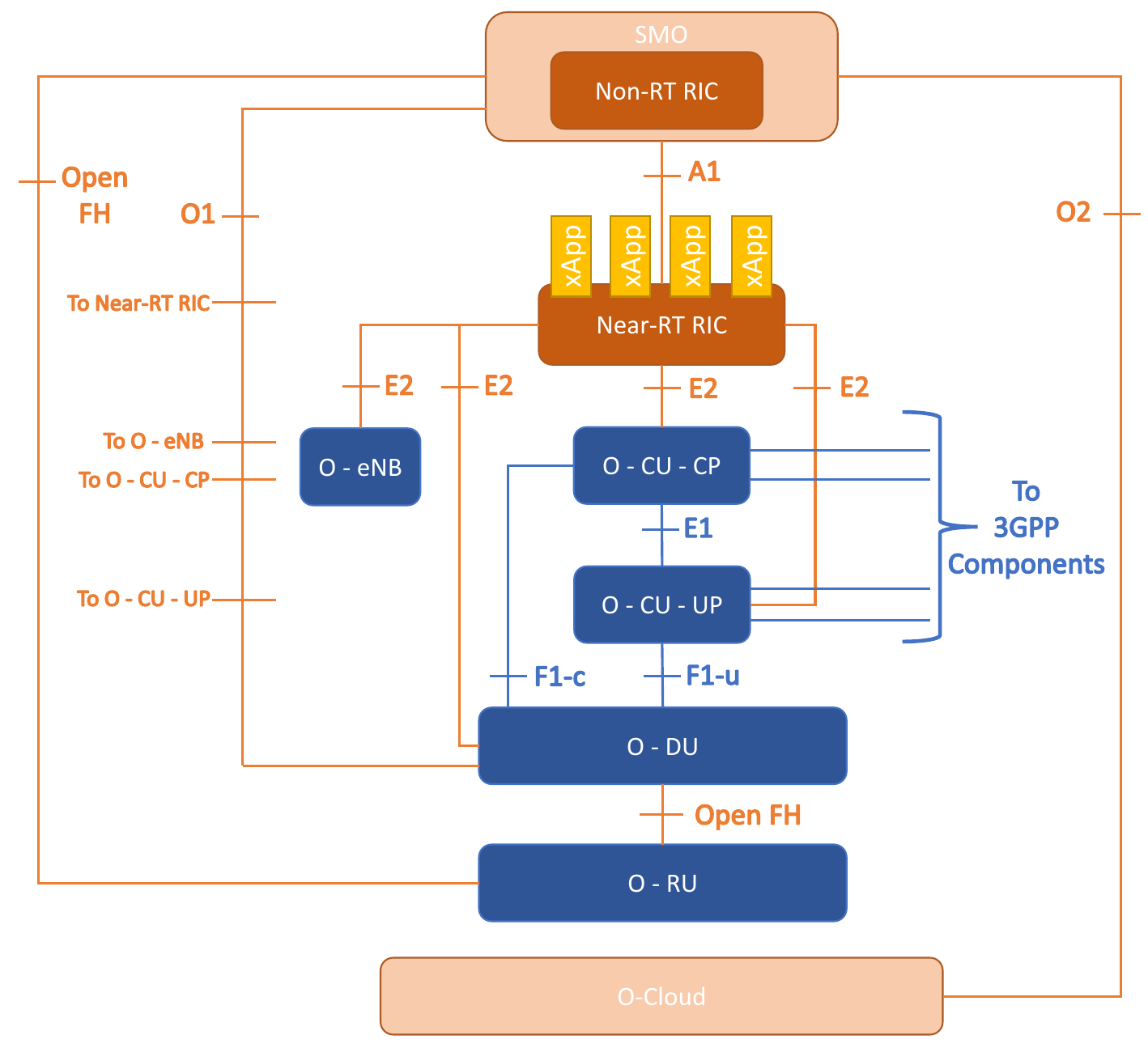}
  \caption{Open Radio Access Network O-RAN Architecture \cite{alliance2020ran}}\label{arch}
   
 \end{figure}

Within the O-RAN framework, RICs and open interfaces play a pivotal role in enabling AI and ML capabilities. 
The near Real-Time RIC (near-RT RIC) connects to RAN elements through the E2 interface, facilitating control loops operating between 10ms and 1s. On the other hand, the non-Real-Time (non-RT RIC) is integrated into Service Management and Orchestration (SMO) frameworks, operating at timescales larger than 1s and connecting to near-RT RIC through the A1 interface. As shown in Fig. \ref{arch}, these RICs adopt a centralised perspective enabled by data pipelines that continuously flow and aggregate Key Performance Measurements (KPMs) involving the network infrastructure's status, such as user count, load, throughput, and resource utilisation, along with additional context information via RICs interfaces with RAN entities. By processing and analysing this worthy information, both RICs employ AI/ML algorithms to implement control policies and actions within the RAN based on the established policies. This paradigm presents data-driven, closed-loop control, enabling automated optimisation of network and load balancing, scheduling policies, handovers, and other essential functionalities \cite{dryjanski2021ran}.

The integration of AI and ML within O-RAN brings numerous advantages. Firstly, these technologies enhance network optimisation by analysing extensive data collected from various network elements. AI/ML algorithms identify patterns and anomalies, predict traffic patterns, and optimise resource allocation for improved efficiency and user experience. Secondly, AI and ML capabilities enable intelligent automation of network operations and routine tasks and minimise human intervention. Furthermore, AI algorithms provide actionable insights and recommendations to network operators, facilitating data-driven decision-making, network quality enhancement, and improved service delivery to end-users. The application of AI and ML facilitates intelligent decision-making, providing network operators with actionable insights and recommendations for improved performance and resource utilisation \cite{polese2023understanding}. 

\subsection{Related Work}
Several studies have explored the potential of AI and ML algorithms in optimising network operations, automating tasks, and enabling intelligent decision-making within the O-RAN framework from both the RAN side and the user side itself \cite{qazzaz2023low}. Here, we highlight some key contributions in this field.
ColO-RAN, a large-scale O-RAN testing framework with software-defined radios-in-the-loop, has been introduced in \cite{polese2022colo}. This framework enables ML research at scale, allowing for the design, training, testing, and evaluation of deep reinforcement learning (DRL)-based closed-loop control in the O-RAN. The framework showcases the importance of experimental frameworks for developing intelligent RAN control pipelines, emphasising the design and testing of DRL agents \cite{polese2022colo}. 

Another area of exploration focuses on utilising O-RAN Fronthaul security due to the highly sensitive data transports between the distributed unit and radio unit. Due to its efficiency, a solution was proposed by Dik and Berger in \cite{dik2021transport} by suggesting MACsec as an effective solution for O-RAN Fronthaul protection. MACsec is a standard security protocol that operates at the data-link layer and provides performance advantages over higher-layer security protocols like IPsec. Additional analysis has been done to ensure its suitability for securing different packet types in the fronthaul. 

Another notable work in the O-RAN field is the development of conflict mitigation algorithms between xAPPs within the architecture. Zhang \emph{et al} proposed in \cite{zhang2022team} a team learning algorithm that is adaptable for multi-vendor RANs, presenting a solution to improve cooperation among xApps. By allowing xAPPs to share their intended actions, this scheme leverages the intentions of other xApps as part of Deep Q-Network (DQN) training and action selection. Furthermore, a conflict mitigation framework (CMF) has been introduced in the O-RAN architecture and integrated into the Conflict Mitigation component of the Near-RT RIC by Adamczyk \emph{et al} in \cite{adamczyk2023conflict}. Their framework detects and resolves diverse conflict types defined in the O-RAN Alliance's specifications, utilising well-defined strategies for detecting each type of conflict, including the exchange of messages between Near-RT RIC components.

Recently, Federated Learning (FL) has emerged as a promising solution for training in disaggregated systems. By facilitating the utilisation of different training inputs for deep reinforced learning (DRL) models, multiple virtual networks can avoid the high cost of collecting the needed data from various RANs in the cellular network \cite{aziz_flcc}. To address this problem, Singh \emph{et al} have proposed in \cite{singh2022mcoranfed} an accelerated gradient descent method and a compression operator to expedite FL convergence and reduce communication costs. They develop a joint optimisation model to select participating trainers in each global round of FL and allocate resources to them while decreasing learning time and resource costs. The proposed FL algorithm (MCORANFed) adheres to the deadline of O-RAN control loops and outperforms state-of-the-art FL methods in terms of convergence and objective costs. In \cite{Rimedotech}, a QoS-based Resource Allocator xApp has been presented to control the percentage of Physical Resource Blocks (PRBs) that should be allocated to the multiple network slices and meet the Service Level Agreements (SLA) requirements while adjusting them depending on the instant traffic demand.

\subsection{Contributions and Problem Statement}
The integration of AI/ML capabilities within O-RAN presents an opportunity to optimise resource allocation, enhance network performance, and meet the diverse QoS requirements of network slices and SLA. However, the efficient allocation of radio resources based on real-time traffic demands poses a significant challenge, including considering factors such as minimum required user throughput, minimum user outage, and delay. Static resource allocation approaches are insufficient as they do not adapt to the varying demands of different network slices. To address this problem, this article presents an ML-based xApp to provide dynamic resource allocation within O-RAN networks. By leveraging the native intelligence of O-RAN controllers: the Near-RT RIC and Non-RT RIC, the proposed algorithm enables proactive actions and auto resource allocation to ensure the optimal utilisation of available radio resources, leading to enhanced network performance and cost-effective delivery of high-quality services.

\section{Dynamic resource allocation in 5G Networks and ML}
\subsection{Resource Allocation in 5G Networks}
Dynamic resource allocation is an important aspect of cellular networks as it enables the efficient utilisation of network resources. The concept of Network Slicing presents an evolutionary approach that exceeds the traditional idea of a singular “best network”. Instead, it defines the creation of multiple logical networks, known as slices, layered upon the physical infrastructure. These slices are designed to cater to different requirements, effectively offering a variety of specialised services to meet a wide range of needs.

\begin{figure}
  \centering
  \includegraphics[width=\linewidth]{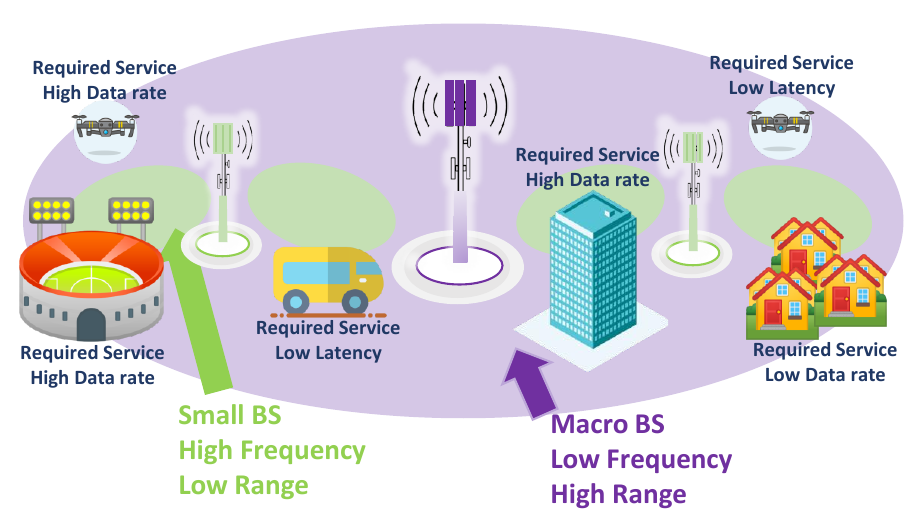}
  \caption{A typical architecture of a 5G network where small and macro cells are deployed to serve multiple user demands}\label{city}
 \end{figure}

In the context of 5G networks, as shown in Figure \ref{city}, a specific feature is the ability to categorise user demands based on QoS flows. Each 5G cell operates with specified radio resources, making it essential to allocate them efficiently to network slices to fulfil Service Level Agreements (SLAs) as well as meet individual QoS flow criteria. A dynamic strategy to tackle this challenge involves splitting the resources in response to real-time traffic demands. This adaptive allocation approach ensures optimal resource utilisation and enhances general network performance.

Splitting the PRBs between the associated users as per their required QoS is done inside the RAN controllers as per O-RAN Alliance specifications, where xApps and rApps are needed to perform multiple policies as per the mobile network operators (MNOs). rApps and xApps are software-specialised applications installed on the non-RT and the near-RT RIC, respectively. These applications serve as software components that boost the functionality of the RICs. A remarkable aspect of these apps is their potential to be developed by third-party vendors, independent of the RIC vendor, as O-RAN encourages a more open and cooperative ecosystem. 

\subsection{Near Real Time RIC Implementation}
The Near-RT RIC, where the xApp is installed, controls the allocation of Physical Resource Blocks (PRBs) that should be allocated to the different network slices to meet their SLA requirements. To optimise the network functionality inside the Near-RT RIC, monitoring the SLA-defined requirements of the networks is mandatory within the accepted time (10ms to 1s). Multiple required parameters need to be observed continuously, such as DL/UL UE throughput, Radio Resource Utilization, DL/UL total available PRB, Number of Active UEs, and other parameters via the E2 interface from the O-RAN central unit (O-CU) and the O-RAN Distributed unit (O-DU). 

\begin{figure}
  \centering
  \includegraphics[width=\linewidth]{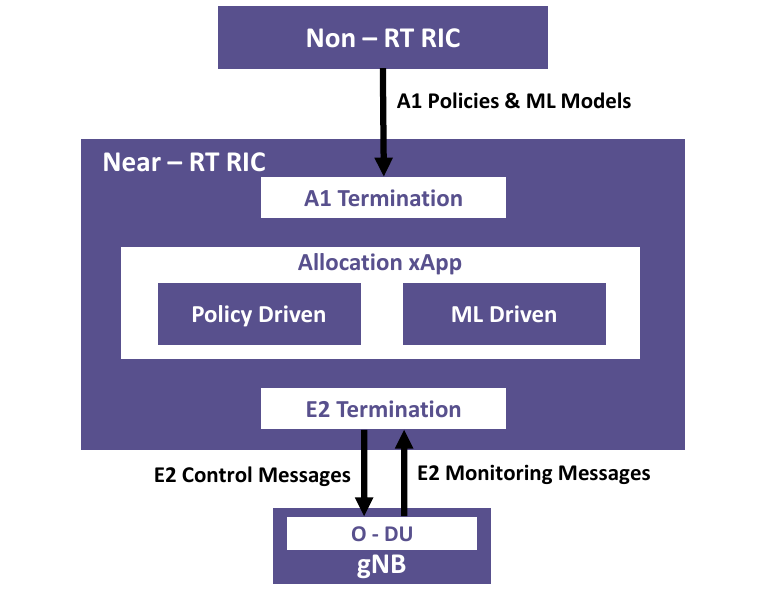}
  \caption{RIC and xApp Deployment}\label{RIC}
 \end{figure}

On the other hand, to enable the Near-RT RIC to obtain full closed-loop control of the O-CU/O-DU via the E2 interface, multiple actions could be taken by changing different network parameters inside CU/DU such as the Min\&Max PRB Policy Ratio and the dedicated PRB Policy Ratio as well as the radio resources splitting policy between users. As shown in Figure \ref{RIC}, the Near-RT RIC sends the E2 commands to the O-DU, where the Medium Access Control (MAC) scheduler operates, to specify the policy for slice management. The policies by which the xApp is controlled are provided through the A1 interface. The A1-P messages are implemented based on O-RAN alliance specifications to comply with the standard. Each policy performs a different way of base station resource allocation when PRBs are allocated so as to provide all the needed slices with the optimal number of PRBs.

The xApp is triggered either by event occasions when the network status changes due to a handover action or configuration change or by a constantly triggered timer which the network administrator could select to monitor the network periodically and solve any emerging issues.

\section{Resource Allocation ML-Based xApp}
\subsection{System Model}
Including ML capabilities within the dynamic resource allocation in O-RAN networks brings significant advantages. ML can play a crucial role in optimising resource allocation by analysing vast amounts of data and identifying complex patterns and correlations.

Based on the provided details, the focus of this article lies in the development and deployment of an ML-based xApp to optimise the allocation of PRBs within each base station and select the most suitable resource allocation policy for the network slices. An ML model must be trained in the SMO framework or directly in the non-RT RIC. The well-trained model can be deployed in the near-RT RIC, while updates and enrichment information of the ML model are transmitted from the non-RT RIC to the near-RT RIC through the A1 interface.

To build the xApp and ML model, we have built a HetNet 5G scenario through extreme programming with Python programming comprising a high-power macro base station covering an area of 20 km × 20 km and a small cell in an area of 10 km × 10 km. Two separated frequency bands, 800 MHz and 2 GHz, with channel bandwidths of 5 MHz and 10 MHz, respectively, are chosen. Voice and Mobile Broadband (MBB) users are randomly deployed within the simulated area, with varying numbers and locations. Multiple resource allocation policies are explored in the near-RT RIC and xApp to determine how available resources should be managed and allocated to the associated users to meet the required data rates of the users.

\definecolor{mycolor}{HTML}{D15FEE}
\begin{mdframed}[backgroundcolor=mycolor!10]
\textbf{Model Parameters}:
\begin{itemize}
    \item Network: 5G Heterogeneous Network (HetNet)
    \item Base Stations: One Macro Base Station (MBS) and One Small Base Station (SBS)
    \item Coverage Area: 20 km × 20 km for MBS and 10 km × 10 km for SBS
    \item Frequency Bands: 800 MHz and 2 GHz
    \item Channel Bandwidth: 5 MHz and 10 MHz
    \item Type of UEs: Voice and Mobile Broadband (MBB)
    \item Minimum QoS Requirement: 250 Kbps for Voice UEs and 3 Gbps for MBB UEs
\end{itemize}
\end{mdframed}

To address the importance of the dynamic resource allocation adjustment, four different resource allocation policies are proposed. These policies are observed through extensive simulations to comprehensively assess their impact on network performance. When distributing the total allocated resources for a base station, which is denoted as $PRBs$, among the associated voice users, represented by $N_{\text{voice}}$, and the associated MBB users, denoted as $N_{\text{mbb}}$, users will collectively utilize these resources, represented by $PRB_{user}$ according to one of the following policies:

\begin{itemize}
\item \emph{Equal Allocation Policy}: Under this policy, an equal split of PRBs between voice and MBB users will occur, disregarding their respective QoS requirements as described in (\ref{1}). This policy is suitable when the number of users is relatively low.
\begin{equation}\label{1}
PRB_{user}=\frac{PRBs}
    {N_{\text{voice}} + N_{\text{mbb}}}
\end{equation}

\item \emph{Voice Priority Allocation Policy}: This policy allocates more resources ($M$ times) to voice users within each cell type to ensure a higher priority for network access. It is an ideal selection for scenarios where voice services are important. The allocated $PRBs$ for each voice user donated by $PRB_{voice}$, and for each MBB user, $PRB_{mbb}$, is represented in (\ref{2})

\begin{gather}
\begin{aligned}\label{2}
PRB_{mbb}=\frac{PRBs}
    {(M*N_{\text{voice}}) + N_{\text{mbb}}} \\
PRB_{voice}=M*PRB_{mbb}
\end{aligned}
\end{gather}

\item \emph{MBB Priority Allocation Policy}: Alternatively, this policy allocates more resources to MBB users ($K$ times) within each cell type compared to voice users as defined in (\ref{3}). This approach prioritises broadband services and is valuable when dealing with high data traffic requirements.

\begin{gather}
\begin{aligned}\label{3}
PRB_{voice}=\frac{PRBs}
    {N_{\text{voice}} + (K*N_{\text{mbb}})} \\
PRB_{mbb}=K*PRB_{voice}
\end{aligned}
\end{gather}

\item \emph{Dedicated Resources Reservation Policy}: This policy reserves a specific portion of the resources for voice users and another portion for MBB users within cells. This ensures dedicated resources for essential services, even during peak network usage and saves the remaining resources for other services. For voice users connected to macro base stations, the resource allocation policy involves dividing a reserved portion, denoted as $\alpha$, from the total available $PRBs$ equally as represented in (\ref{4}). On the other hand, voice users connected to small base stations will also receive an equivalent allocation from the voice users connected to macro base stations. The allocation of resources for MBB users, which is defined in (\ref{5}), follows a different approach. These resources are distributed based on a selected portion, denoted as $\beta$, from the total $PRBs$ regardless of whether they are connected to macro or small base stations.

In macro base stations
\begin{equation}\label{4}
PRB_{voice}=\frac{PRBs}
    {\alpha * N_{\text{voice}}}
\end{equation}
And in any kind of base stations
\begin{equation}\label{5}
PRB_{mbb}=\frac{PRBs}
    {\beta * N_{\text{mbb}}}
\end{equation}

\end{itemize}

\subsection{Results and Discussion}
The performance and effectiveness of the mentioned resource allocation policies, in conjunction with the ML model, have been thoroughly investigated and evaluated. We investigate the impact of each resource allocation policy on network performance, taking into account a diverse deployment of associated users with different types. To evaluate the effect of different policies on network performance, the network outage, a critical performance metric, has been considerably analysed. The measurement of network outage was achieved by determining the percentage of users who failed to get the minimum QoS requirements out of the total number of the associated users within the cell, disregarding the user types. Multiple simulations have been conducted, encompassing various user deployment scenarios within the network, to accurately capture the system outage for each configuration and policy setting.

\begin{figure}
  \centering
  \includegraphics[width=\linewidth]{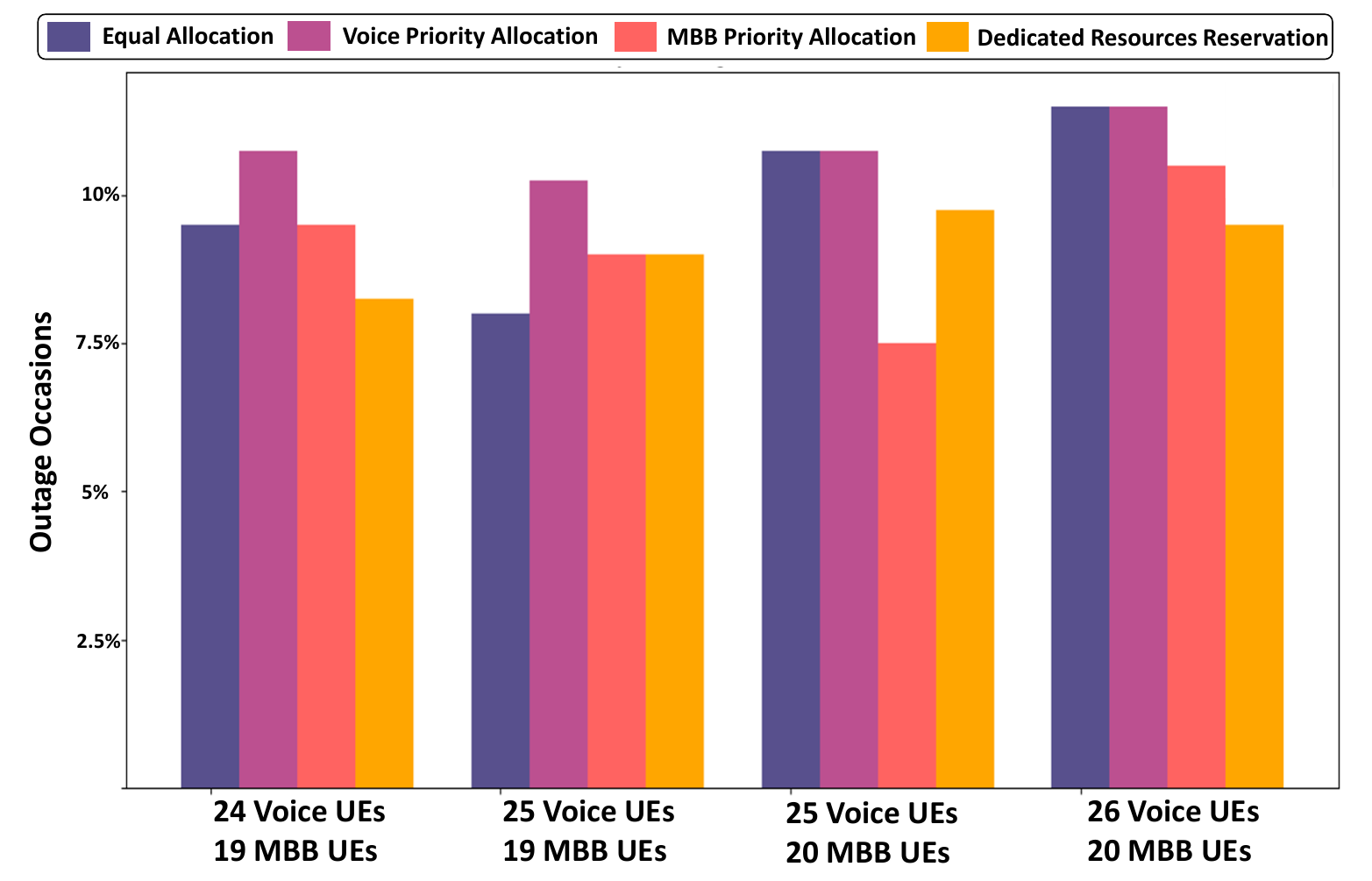}
  \caption{Small Cell Network Outage under Different Policies}\label{small_outage_2d}
 \end{figure}

The simulations produced valuable insights into the behaviour of our 5G simulated network. To emphasise the variance in network performance under different resource allocation policies, we considered high QoS requirements for each user in our simulation. We observed that, even under similar conditions concerning user types and numbers, the implementation of each of the four available policies led to varying outage events.
Figure \ref{small_outage_2d} presents the outage occurrences when different policies are applied to particular network configurations in a small cell. In particular, the Dedicated Resources Reservation policy yielded the least user outages when there were 24 voice users and 19 MBB users. However, when an additional voice user is associated with the same cell, implementing the same policy does not yield optimal performance, whilst the Equal Allocation policy is shown to be more effective. Similar observations in performance were noticed when cells had more users of different types. These impacts emphasise the need for dynamic selection of policy based on the network conditions and the number and types of users associated with each cell.

Accordingly, out of that, we proceeded to develop an ML model trained using the comprehensive and detailed information collected from the network to select the optimal performance policy for each network condition reasonably. We aim to enhance the overall act of resource allocation in 5G networks and lead to improving network performance, user satisfaction, and overall Quality of Experience (QoE).

To train the ML model, comprehensive simulations were conducted to generate datasets consisting of a large number of input pairs. These input pairs included the count of the users from both voice and MBB user types, along with corresponding outage information and network configuration. The objective of the classification problem is to determine the best allocation policy for each network configuration, as certain policies may result in suboptimal network performance.

\begin{figure}
  \centering
  \includegraphics[width=\linewidth]{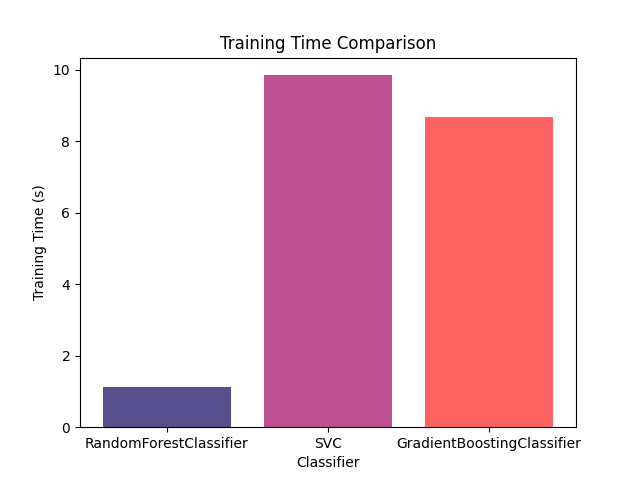}
  \caption{ML Training Time}\label{ML2}
 \end{figure}

After careful evaluation of various classification algorithms, multiple algorithms show an excellent performance in terms of classification accuracy during the model validation, i.e. more than 85\%. However, some of the algorithms, as demonstrated in Figure \ref{ML2}, indicate a different training period which might be considered as an essential factor, especially in the case of repeating the training periodically or performing the online training event-based. That is why the Random Forest classifier was chosen as the ML algorithm for this study.

The utilisation of the Random Forest classifier as the ML algorithm, along with the dataset collection and simulations, contributes to the effectiveness and reliability of the proposed resource allocation framework.

\begin{figure}
  \centering
  \includegraphics[width=\linewidth]{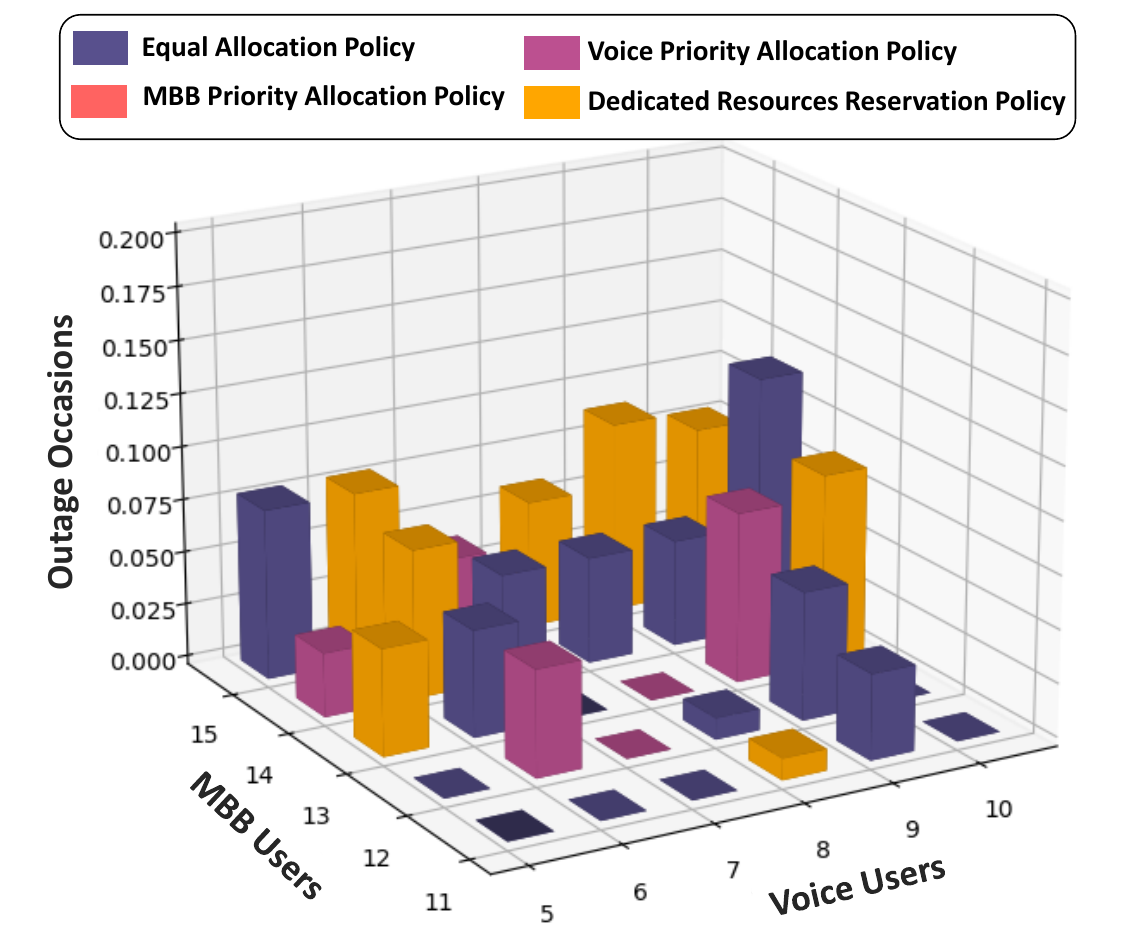}
  \caption{Macro Cell Network Outage }\label{macro_outage}
 \end{figure}

Appling our trained model presents a dynamic allocation mechanism that can intelligently select the most suitable policy based on the specific configuration and user demands. Fig.\ref{macro_outage} illustrates the selected policy for each network configuration and user count. A slight variation in the number of associated users from different user types can significantly impact the occurrence of network outages. Choosing the best performance policy will ensure the minimum outage at each cell state.   

Furthermore, it is worth mentioning that within the macro cell observations, the MBB Priority Allocation policy rarely emerged as the best allocation choice, while a similar remark was observed with the Voice Priority Allocation policy in the small cells. In the case of the MBB Priority Allocation policy, it ensures that all cells, regardless of their kind, allocate resources primarily to MBB users, no matter the number of connected users. While this approach guarantees a prioritised service for MBB users, it may lead to a higher outage for voice users and system outages at the end. The same action happens with the Voice Priority Allocation policy within the small cell results, where using it leads to higher outages than using the other available policies.
 
The integration of ML within the xApp enables a responsive and efficient resource allocation process. By accurately estimating the current traffic demands and dynamically adapting the allocation of PRBs, ML algorithms facilitate the efficient utilisation of network resources and enhance the overall network performance. This approach ensures that the network slices receive the appropriate allocation of resources, leading to improved user experience, optimised network efficiency, and enhanced fulfilment of SLAs and QoS requirements. By leveraging historical data and real-time monitoring, ML algorithms optimise the allocation of PRBs to meet the dynamic traffic demands and ensure the best possible network performance. The utilisation of ML algorithms within the xApp empowers O-RAN networks to achieve efficient and intelligent resource management, contributing to enhanced user experience and network efficiency.

\section{Conclusion}
In conclusion, the integration of AI/ML capabilities within the O-RAN  brings considerable advantages to the telecommunications industry. This paper presents an xApp-based implementation for dynamic resource allocation using ML within O-RAN architecture. A Het-Net 5G network has been built to explore the consequences and observations of resource allocation policies. Macro and small 5G cells deployed with different frequency bands and bandwidths. Four different policies have been implemented at multiple network configurations, and the user outage was chosen as a performance metric. Our simulation experiments show that
under similar network conditions, the implementation of each of the four policies led to different outage events. Therefore, an ML-based xApp has been developed to enable a dynamic selection for resource allocation policies. The proposed ML model effectively picks the best allocation policy for each base station, optimising resource allocation and achieving Service Level Specifications (SLS). With a high accuracy rate of 85\% for policy classification, the Random Forest Classifier displays its potential to enhance scheduler functionality within the O-DU. Leveraging input parameters from the O-RAN instance conditions, the xApp performs rapid convergence to changing network conditions. This study highlights the transformative capability of ML in advancing network efficiency and performance, paving the way for more resilient and adaptable telecommunication networks.

\section*{Acknowledgment}
This research was funded by EP/X040518/1 EPSRC CHEDDAR and was partly funded by UKRI Grant EP/X039161/1 and MSCA Horizon EU Grant 101086218. 

\bibliographystyle{unsrt} 
\bibliography{ref} 

\begin{thebibliography}{10}

\bibitem{bonati2020open}
Leonardo Bonati, Michele Polese, Salvatore D’Oro, Stefano Basagni, and Tommaso Melodia.
\newblock Open, programmable, and virtualized 5g networks: State-of-the-art and the road ahead.
\newblock {\em Computer Networks}, 182:107516, 2020.

\bibitem{masur2022artificial}
Paul~H Masur, Jeffrey~H Reed, and Nishith~K Tripathi.
\newblock Artificial intelligence in open-radio access network.
\newblock {\em IEEE Aerospace and Electronic Systems Magazine}, 37(9):6--15, 2022.

\bibitem{plantin2021geopolitical}
Jean-Christophe Plantin.
\newblock The geopolitical hijacking of open networking: the case of open ran.
\newblock {\em European Journal of Communication}, 36(4):404--417, 2021.

\bibitem{wypior2022open}
Dariusz Wypi{\'o}r, Miros{\l}aw Klinkowski, and Igor Michalski.
\newblock Open ran—radio access network evolution, benefits and market trends.
\newblock {\em Applied Sciences}, 12(1):408, 2022.

\bibitem{bonati2021intelligence}
Leonardo Bonati, Salvatore D'Oro, Michele Polese, Stefano Basagni, and Tommaso Melodia.
\newblock Intelligence and learning in o-ran for data-driven nextg cellular networks.
\newblock {\em IEEE Communications Magazine}, 59(10):21--27, 2021.

\bibitem{dryjanski2021ran}
M~Dryjanski and R~Lundberg.
\newblock The o-ran whitepaper; overview, architecture, and traffic steering use case.
\newblock {\em Overview Architecture and Traffic Steering Use Case}, 2021.

\bibitem{alliance2020ran}
Open~RAN Alliance.
\newblock O-ran-wg1-o-ran architecture description-v01. 00.00.
\newblock {\em Technical Specification}, 2020.

\bibitem{polese2023understanding}
Michele Polese, Leonardo Bonati, Salvatore D’oro, Stefano Basagni, and Tommaso Melodia.
\newblock Understanding o-ran: Architecture, interfaces, algorithms, security, and research challenges.
\newblock {\em IEEE Communications Surveys \& Tutorials}, 2023.

\bibitem{qazzaz2023low}
Mohammed M.~H. Qazzaz, Syed~A Zaidi, Des McLernon, Abdelaziz Salama, and Aubida~A Al-Hameed.
\newblock Low complexity online rl enabled uav trajectory planning considering connectivity and obstacle avoidance constraints.
\newblock In {\em 2023 IEEE International Black Sea Conference on Communications and Networking (BlackSeaCom)}, pages 82--89. IEEE, 2023.

\bibitem{polese2022colo}
Michele Polese, Leonardo Bonati, Salvatore D’Oro, Stefano Basagni, and Tommaso Melodia.
\newblock Colo-ran: Developing machine learning-based xapps for open ran closed-loop control on programmable experimental platforms.
\newblock {\em IEEE Transactions on Mobile Computing}, 2022.

\bibitem{dik2021transport}
Daniel Dik and Michael~St{\"u}bert Berger.
\newblock Transport security considerations for the open-ran fronthaul.
\newblock In {\em 2021 IEEE 4th 5G World Forum (5GWF)}, pages 253--258. IEEE, 2021.

\bibitem{zhang2022team}
Han Zhang, Hao Zhou, and Melike Erol-Kantarci.
\newblock Team learning-based resource allocation for open radio access network (o-ran).
\newblock In {\em ICC 2022-IEEE International Conference on Communications}, pages 4938--4943. IEEE, 2022.

\bibitem{adamczyk2023conflict}
Cezary Adamczyk and Adrian Kliks.
\newblock Conflict mitigation framework and conflict detection in o-ran near-rt ric.
\newblock {\em IEEE Communications Magazine}, 2023.

\bibitem{aziz_flcc}
Abdelaziz Salama, Syed~Ali Zaidi, Des McLernon, and Mohammed M.~H. Qazzaz.
\newblock Flcc: Efficient distributed federated learning on iomt over csma/ca.
\newblock In {\em 2023 IEEE 97th Vehicular Technology Conference (VTC2023-Spring)}, pages 1--6, 2023.

\bibitem{singh2022mcoranfed}
Amardip~Kumar Singh and Kim~Khoa Nguyen.
\newblock Mcoranfed: Communication efficient federated learning in open ran.
\newblock In {\em 2022 14th IFIP Wireless and Mobile Networking Conference (WMNC)}, pages 15--22. IEEE, 2022.

\bibitem{Rimedotech}
Rimedo Labs.
\newblock {Quality of Service-based Resource Allocator xApp (QRA-xApp)}.
\newblock Technical specification, {RIMEDO Labs}, 2022.

\end{thebibliography}

\vspace{12pt}

\end{document}